\documentclass[prb,twocolumn,amssymb,showpacs]{revtex4}
\usepackage{graphicx}
\usepackage{amsmath}

\begin{document}

\title{Magnetic phase transition in CoO confined to a vycor type porous glass. Neutron diffraction study.}
\author{I. V. Golosovsky}
\affiliation{Petersburg Nuclear Physics Institute, 188300, Gatchina, St. Petersburg, Russia.}
\author{I. Mirebeau, G. Andr\'e}
\affiliation{Laboratoire L\'eon Brillouin, CE-Saclay, F-91191, Gif-sur-Yvette, France.}
\author{D. A. Kurdyukov and Y. A. Kumzerov}
\affiliation{A. F. Ioffe Physico-Technical Institute, 194021, St. Petersburg, Russia.}
\author{M. Tovar, D. M. T\"{o}bbens,}
\affiliation{Hahn-Meitner-Institut, D-14109 Berlin, Germany.}

\begin{abstract}
Neutron diffraction studies of antiferromagnetic CoO confined to a vycor type porous glass demonstrate a
continuous magnetic phase transition with a decreased N\'eel temperature and reduced magnetic moment as
compared with the bulk.
\end{abstract}

 \maketitle

\section{Introduction}

The behavior of the nanostructured substances in the conditions of so-called "restricted" or
"confined" geometry has attracted great fundamental and practical interest because of their potential use for
technological application. One of the most interesting problem is magnetism in confinement.

Oxides of the transition 3d-metals as MnO and CoO are especially interesting for such investigations. They
have a similar antiferromagnetic structure. In MnO the magnetic moments lay within the (111) plane, while in
CoO they tilted out from this plane \cite{Shull,Roth,Herrmann}. It leads to different magnetic symmetry that
results in different magnetic behavior: in MnO the magnetic order appears by a first order transition while
in CoO it does by a second order transition \cite{Rechtin,Mukamel}.

The magnetic transition in MnO is accompanied by a rhombohedral crystal distortion which stabilizes an
antiferromagnetic structure. In CoO one observes a strong tetragonal distortion accompanied by a small
trigonal distortion \cite{Jauch}. At last in MnO the orbital magnetic moment is practically "frozen", while
in CoO the large orbital contribution is observed \cite{Neubeck,Flipse}

After the neutron diffraction experiments with antiferromagnet MnO embedded in a porous glass \cite{MnO-PRL}
we performed the experiments with CoO synthesized within the similar porous matrix. Our objective was to
compare the behavior of these oxides in the conditions of "restricted" geometry.

\section{Experiment }

As well as in the case of MnO the host porous matrix was made from a sodium borosilicate glass \cite{Levitz}.
The matrix has a random interconnected network of elongated pores \cite{Pb} with a narrow distribution of
pore diameters of 70(3) {\AA}. CoO was synthesized from a cobalt nitrate solution by a chemical, "bath
deposition" method.

The chemical instability of CoO in the conditions of high dispersed nanoparticles with large ratio
surface/volume strongly increases. Therefore without special precautions CoO oxidizes after about a year in
practically stoichiometric Co$_3$O$_4$.

Because of small quantity of CoO embedded within nanopores the signal is weak. Therefore for measurement of
magnetic reflections at low angles of diffraction we used the high luminosity diffractometer G6-1 of the
Laboratoire L\'eon Brillouin with a large neutron wavelength of 4.73 {\AA}. For measurements of nuclear
reflections we used diffractometers with better resolution and smaller neutron wavelength G4-1 of the
Laboratoire L\'eon Brillouin and E9 of the Hahn-Meitner-Institut with neutron wavelengths 2.43 {\AA} and 1.80
{\AA}, respectively.

\section{Results and discussion}

\begin{figure} [t]
\includegraphics* [width=\columnwidth] {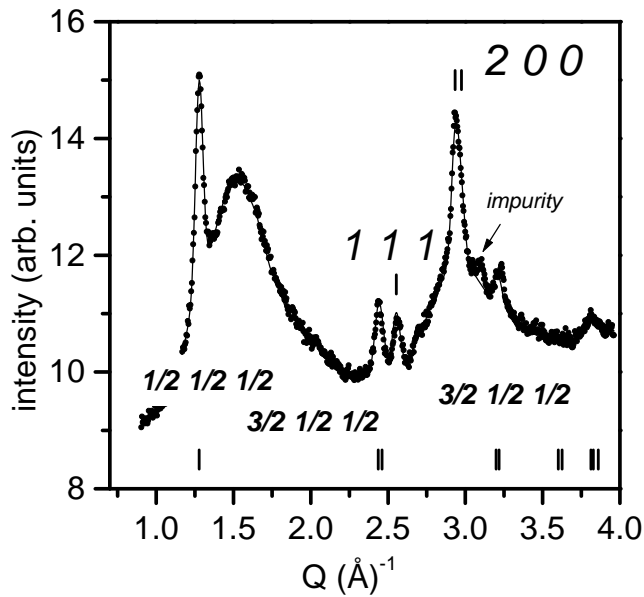}
\caption{ Neutron diffraction patterns from CoO confined within a porous glass, observed (closed circles) and
calculated (solid line). Diffractometer G4-1, temperature 10 K.} \label{profile}
\end{figure}

Neutron diffraction patterns (figure \ref{profile}) show the crystallizing of the CoO within nanopores. The
amount of the impurity phase Co$_3$O$_4$ (its reflection is shown by an arrow in figure \ref{profile}) does
not exceed some percents.

From the peak broadening the volume-averaged diameter of the embedded nanoparticles was estimated as 100(5)
{\AA}. This value is larger than the mean pore diameter 70 {\AA}, showing that the confined oxide CoO,
similar MnO, forms interconnected aggregates rather than isolated nanoparticles. It demonstrates that the
crystallization spreads over at least several adjacent pores. The measured averaged diameter turns out to be
smaller than the nanoparticle diameter of 140 {\AA} found in the embedded MnO, that is connected with
different wetting ability of two oxides.

The indexing of the observed magnetic reflections below 280 K corresponds to antiferromagnetic ordering of
type-II in the fcc lattice similar to the bulk CoO \cite{Shull,Roth}. The collinear magnetic moments were
found tipped out of the (111) plane $9.5(3)^o$  that is consistent with the reported value for the bulk
\cite{Herrmann}.

From the magnetic Bragg reflections, the ordered magnetic moment at 10 K was found to be 2.92(2) $\mu_B$/ion.
This volume-averaged value is smaller than the value of 3.80 (1) $\mu_B$/ion measured for the bulk CoO
\cite{Herrmann}. The similar phenomenon was observed for MnO embedded within different porous media and was
attributed to the disordering of moments at the surface of nanoparticle \cite{MnO-PRL, MnO-channels}.

The observed structure distortions was found to be the same as in the bulk, namely, at the lowest temperature
the tetragonal distortion (a-c)/c, (here \emph{a} and \emph{c} are lattice parameters) in confined CoO is
0.0114(7) that well coincides with that measured for the bulk\cite{Jauch}. Small trigonal distortion observed
in the synchrotron radiation experiments is 100 times weaker than tetragonal distortion and is beyond of the
accuracy of our experiment \cite{Jauch}.

In spite of rich history of experimental and theoretical studies of the bulk CoO the data about its critical
behavior are controversial. This oxide is traditionally considered as a classic example of compound with a
large unquenched orbital moment. Therefore its magnetic behavior should follow to the prediction of the
3D-Ising model with critical exponent $\beta=0.312$ (or 5/16), that was shown experimentally in the
specific heat measurements \cite{Salomon}.

The birefringence measurements of \cite{Germann} showed the close value of $\beta = 0.29(2)$. However, the
neutron diffraction with a single crystal displayed $\beta  = 0.25(2)$ only \cite{Rechtin}. To explain this
disagreement the phenomenological correction for the tetragonal lattice contraction below $T_N$ was proposed,
which raises $\beta$ to values of 0.29(2) \cite{Rechtin}.

Moreover, in ref.\cite{Jauch} the authors found upon the symmetry breaking from cubic to monoclinic symmetry
in the bulk CoO concluded that the magnetic transition is discontinuous but with very small first order jump.

\begin{figure} [t]
\includegraphics* [width=\columnwidth] {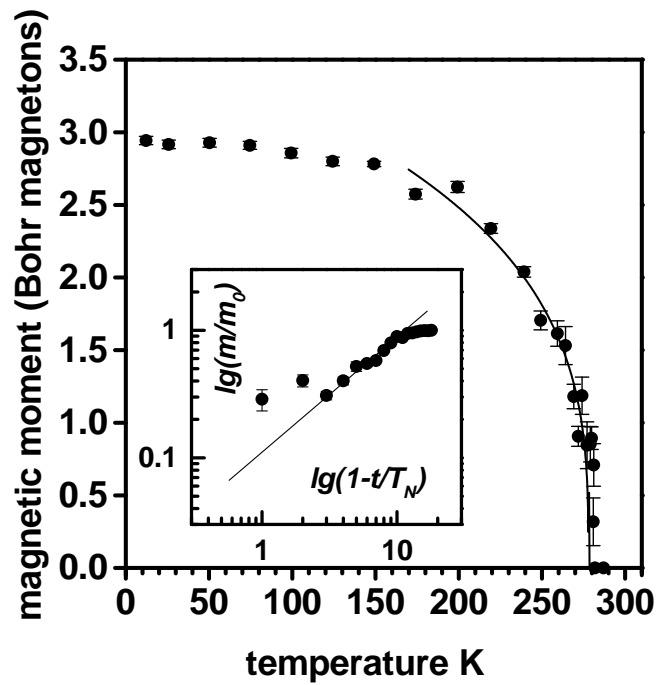}
\caption{Temperature dependence of the magnetic moment and its fitting with a power law. In the inset the
log-log graph is shown. It is seen that two points nearest to $T_N$ have a large contribution from
quasi-elastic scattering} \label{moment}
\end{figure}

In figure \ref{moment} the temperature dependence of the magnetic moment for confined CoO is shown. Fitting
the observed dependence with a power law $m(T) = m_0(1-T/T_N)^\beta$ yield the value $\beta = 0.31(2)$. This
value of $\beta$ should be compared with the uncorrected for contraction value of 0.25(2) measured for the
bulk \cite{Rechtin}. The increasing of the critical exponent in confinement one can explain by the well known
effect of the finite-size rounding of the phase transition\cite{Ymry}, which leads to a smearing up the
transition that, in turn, goes to increasing of the critical exponent.

Surprisingly, the measured value of $\beta$ in confinement appears to be close to the expected value for the
bulk 3D-Ising model, however it is smaller than the critical exponent of 0.362(4) calculated by a computer
simulation of the finite-size scaling for 3D-Ising model \cite{Landau}.

N\'eel temperature  $T_N = 278(0.5)$ K turns out to be decreased as compared with the bulk value of 289.0(1)
K \cite{Rechtin}. It is a common effect expected for nanostructured magnets \cite{Swift,Ambrose}, resulted
from the limiting of the correlation length by the nanoparticle size. However in MnO confined in the same
porous media $T_N$ was found enhanced with respect to the bulk \cite{MnO-PRL,MnO-channels}.

There are some competitive factors which influence the transition temperature. The decreasing the correlation
length in nanostructured objects leads to a decrease of $T_N$. However there is another mechanism of the
ternary interactions of the structural and magnetic order parameters which can lead to an increase of $T_N$
\cite{MnO-channels}. Apparently the relative role of these two factors is different for two oxides CoO and
MnO.

In conclusion, by using neutron diffraction the magnetic order was observed for antiferromagnet CoO embedded
in a porous glass. The type of magnetic order and structural distortion in nanostructured CoO was found to be
the same as in the bulk. The ordered magnetic moment is noticeably smaller than in the bulk. The magnetic
phase transition was found to be continuous as in the bulk, however with decreased $T_N$ and slightly
increased critical exponent that attributed to size-effect rounding of a phase transition in confinement.

The work was supported by the Grants RFBR 04-02-16550, SS-1671-2003.2 and the INTAS No. 2001-0826.

\end{document}